\documentclass[conference]{IEEEtran}
\IEEEoverridecommandlockouts
\usepackage{cite}
\usepackage{amsmath,amssymb,amsfonts}
\usepackage{algorithmic}
\usepackage{graphicx}
\usepackage{textcomp}
\usepackage{xcolor}
\def\BibTeX{{\rm B\kern-.05em{\sc i\kern-.025em b}\kern-.08em
    T\kern-.1667em\lower.7ex\hbox{E}\kern-.125emX}}
\begin{document}

\title{A RFID BASED CAMPUS-WIDE PAYMENT SYSTEM\\
{\footnotesize \textsuperscript{}}
\thanks{}
}

\author{\IEEEauthorblockN{Miraj Uddin Chowdhury}
\IEEEauthorblockA{\textit{Department of Computer Science and Engineering} \\
\textit{International Islamic University Chittagong}\\
Chittagong \\
c201074@ugrad.iiuc.ac.bd}
\and
\IEEEauthorblockN{MD Khairul Islam Prime}
\IEEEauthorblockA{\textit{Department of Computer Science and Engineering} \\
\textit{International Islamic University Chittagong}\\
Chittagong \\
c201100@ugrad.iiuc.ac.bd}
}
\maketitle

\begin{abstract}
This work titled "RFID Based Campus-Wide Payment System" introduces an innovative cashless payment solution for educational institutions. It uses RFID cards and a Raspberry Pi to enable hassle-free payments for various campus services, such as cafeteria purchases, tuition fees, and library fines. A centralized database ensures real-time updates on transactions and account balances, accessible through a simple and user-friendly web interface. This work is involved in designing a secure system with object-oriented principles, setting up databases, and integrating hardware like RFID readers with a Raspberry Pi. The systems are proved to be a cost-effective and efficient alternative to traditional payment methods, enhancing convenience and security for students and administrators. The study also explored similar RFID applications, like smart parking and attendance systems, to identify challenges and improvements. Looking ahead, it envisions features like wearable RFID devices, voice-activated payments, and blockchain integration to boost security and usability. Results show that this system simplifies campus payments and has the potential for broader adoption in similar environments.
\end{abstract}

\section{Introduction}
The rapid advancement of modern technology is transforming traditional payment systems. While paper money has long been the global medium of exchange, it is increasingly being replaced by digital alternatives such as magnetic cards and automated teller machine (ATM) cards. These technologies enhance convenience by eliminating the need for cash-based transactions. Radio-Frequency Identification (RFID) technology further advances this trend by leveraging electromagnetic fields to automatically identify and track objects. An RFID system typically comprises a tiny radio transponder, a receiver, and a transmitter, enabling seamless and secure data transmission.

In the context of academic institutions, traditional cash-based payment systems present significant challenges, including inefficiency, inconvenience, and security risks. Students must carry cash for various purposes, such as tuition payments, canteen expenses, library fines, and admission fees, creating risks of theft and loss. Moreover, cash-based transactions often lead to delays and administrative inefficiencies.

To address these issues, this work proposes an RFID-based campus-wide cashless payment system. This system aims to provide students with a secure, efficient, and convenient payment solution using RFID cards. Such a system would streamline campus financial operations, enhance security, and improve user experience. By eliminating cash handling, the proposed solution offers increased transparency and accountability, benefiting both students and campus administrators.

The design and implementation of this system focus on achieving seamless integration with various campus functionalities, ensuring robust performance and user satisfaction. This paper outlines the system architecture, development methodology, and performance evaluation, highlighting its potential to revolutionize financial transactions within tertiary educational institutions.

\section{Literature Review}

Radio-Frequency Identification (RFID) technology has become increasingly prevalent across various industries due to its ability to facilitate automated, contactless identification through radio signals. Gaikwad et al. (2017) highlighted its potential in cashless systems to provide secure and convenient payment methods. In India, RFID is already being utilized in toll systems to reduce congestion and streamline transactions, demonstrating its effectiveness in addressing inefficiencies (Mall and Shaikh, 2017). The continuous evolution of RFID technology is expected to contribute significantly to various societal and industrial advancements (Duroc, 2018).

Several studies illustrate the diverse applications of RFID technology. For instance, an IoT-based smart parking system integrates RFID cards and IR sensors to minimize waiting times and operational costs while promoting sustainability through reduced paper usage. This system automates parking billing using RFID-tagged vehicles, enhancing both efficiency and revenue generation for facility owners. Similarly, RFID-based attendance systems, as developed by Manav Rachna International University, simplify attendance tracking in educational and professional environments, offering real-time monitoring and enhanced security.

Another notable application is the RFID-based child security system, which combines RFID, GPS, and GPRS technologies to monitor student transportation. This system enhances safety by tracking students’ movements and providing real-time updates to parents and schools, thereby addressing critical safety concerns (Shabaan et al., 2013). RFID-based cashless canteen payment systems further emphasize the technology's potential to simplify and automate transactions, enhancing customer convenience in daily operations.

In comparing RFID with Near Field Communication (NFC), RFID stands out for its longer range and faster data transfer capabilities, making it suitable for one-way communication in large-scale applications. In contrast, NFC is more suited for short-range, secure, and bidirectional communication scenarios, such as mobile payments.

Despite its advantages, RFID technology faces challenges such as reader and tag collisions. Reader collision occurs when signals from multiple RFID readers interfere, while tag collision arises from simultaneous data transmission by multiple tags in a single field. Anti-collision protocols and sequential data collection mechanisms have been developed to address these issues, ensuring the accuracy and efficiency of RFID systems.

The literature demonstrates that RFID technology is highly versatile, offering innovative solutions across multiple domains. These studies provide valuable insights and a foundation for the development of the proposed RFID-based campus-wide payment system, which aims to streamline financial transactions in an academic setting.

\section{Methodology}

The proposed RFID-based campus-wide payment system aims to enhance the efficiency, convenience, and security of financial transactions in educational institutions. The methodology focuses on developing an integrated system comprising RFID cards, RFID readers, and a backend infrastructure. This section describes the design, implementation steps, and operational flow of the system.

\subsection{System Components}
\begin{itemize}
    \item \textbf{RFID Cards:} Small electronic devices containing unique identifiers and antennas for data transmission. These cards enable seamless identification and transaction processing.
    \item \textbf{RFID Readers:} Devices that generate electromagnetic fields to communicate with RFID cards, extracting their unique identifiers and associated data.
    \item \textbf{Backend System:} Serves as the control hub, handling data reception, processing, storage, and management. Key backend functionalities include:
    \begin{itemize}
        \item Receiving and organizing data from RFID readers.
        \item Generating reports and alerts.
        \item Storing transaction records in a structured database.
        \item Facilitating efficient management of campus services, such as libraries, canteens, and administration.
    \end{itemize}
\end{itemize}

\subsection{System Architecture}
\begin{figure}[h]
    \centering
    \includegraphics[width=\columnwidth]{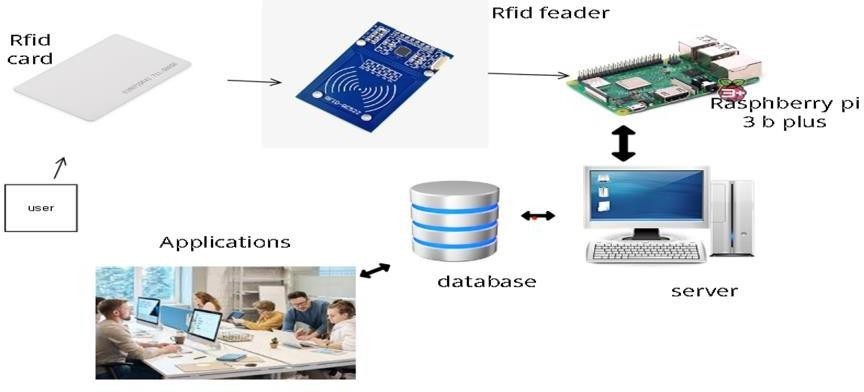}
    \caption{System Architecture of RFID-Based Payment System.}
    \label{fig:system_architecture}
\end{figure}

The architecture integrates RFID hardware with a centralized database and application software. Figure~\ref{fig:system_architecture} illustrates the system architecture, while Table~\ref{table:database} summarizes the core database tables used for various functionalities.

\subsection{Implementation Steps}
\begin{enumerate}
    \item \textbf{Requirement Identification:} Determine the objectives and scope, including integration with library fines, canteen payments, and tuition fees.
    \item \textbf{Selection of Hardware:} Choose appropriate RFID cards and readers compatible with the system's operational environment.
    \item \textbf{Infrastructure Development:}
    \begin{itemize}
        \item Install RFID readers at designated points.
        \item Configure the backend database using tables for students, transactions, and services.
        \item Develop application software for data management and reporting.
    \end{itemize}
    \item \textbf{Testing:}
    \begin{itemize}
        \item Validate system functionality, including card detection, data processing, and payment transactions.
        \item Ensure compatibility with existing campus systems.
    \end{itemize}
    \item \textbf{User Training:} Train administrators and students on system usage and troubleshooting.
    \item \textbf{Deployment:} Implement the system and provide ongoing maintenance for optimal performance.
\end{enumerate}

\subsection{System Workflow}
\begin{figure}[h]
    \centering
    \includegraphics[width=\columnwidth]{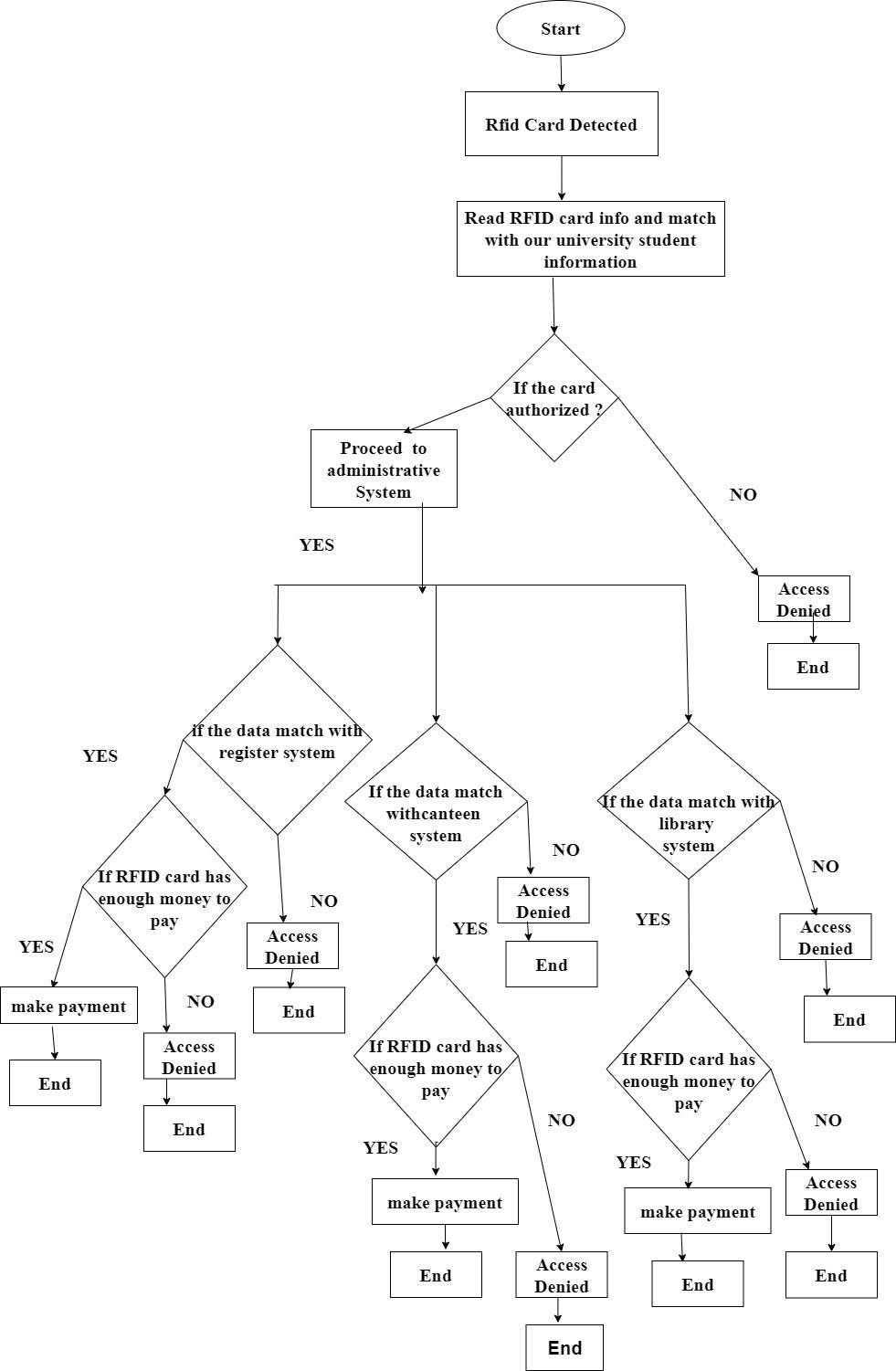}
    \caption{System Workflow of RFID-Based Payment System.}
    \label{fig:system_workflow}
\end{figure}

The operational flow of the system is depicted in the flowchart in Figure~\ref{fig:system_workflow}. Key steps include:
\begin{enumerate}
    \item Detecting the RFID card.
    \item Verifying card authorization.
    \item Checking balance availability.
    \item Completing the transaction or denying access based on system rules.
\end{enumerate}

\subsection{Database Design}
The system utilizes multiple tables to manage data. Table~\ref{table:database} outlines the key database components:

\begin{table}[h]
    \centering
    \caption{Core Database Tables.}
    \label{table:database}
    \begin{tabular}{|p{2cm}|p{2cm}|p{2cm}|}
        \hline
        \textbf{Table Name} & \textbf{Description} & \textbf{Key Fields} \\ \hline
        Student Table & Stores student information. & ID, Name, Email, RFID ID, Balance \\ \hline
        Admin Table & Holds admin details for system control. & ID, Name, Email, Admin Type \\ \hline
        Transaction Table & Logs all financial transactions. & ID, Student ID, Amount, Date \\ \hline
        Recharge History Table & Tracks balance top-ups. & ID, Student ID, Amount, Date \\ \hline
    \end{tabular}
\end{table}

\subsection{Data Flow and Entity Relationships}
The system's data flow and relationships are modeled using Data Flow Diagrams (DFDs) and Entity Relationship (ER) diagrams:
\begin{itemize}
    \item \textbf{DFD Level 0:} Shows the high-level interaction between students, services, and the backend system.
    \item \textbf{DFD Level 1 and 2:} Detail the flow of data between system components, such as the registrar, library, and canteen.
    \item \textbf{ER Diagram:} Represents the relationships between database entities, including students, transactions, and administrators.
\end{itemize}

\begin{figure}[h]
    \centering
    \includegraphics[width=\columnwidth]{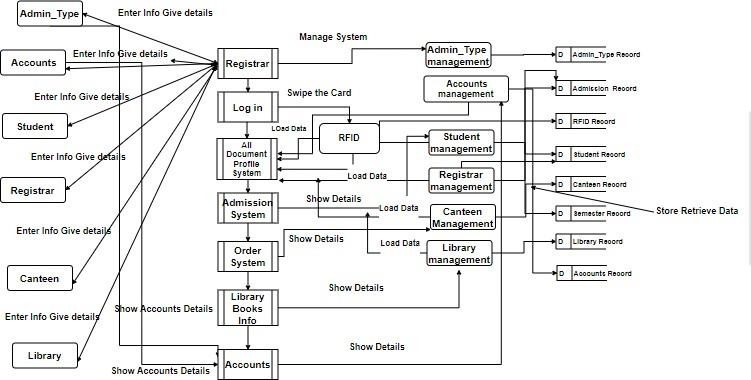}
    \caption{Data Flow Diagram of the RFID-Based Payment System.}
    \label{fig:data_flow}
\end{figure}

\begin{figure}[h]
    \centering
    \includegraphics[width=\columnwidth]{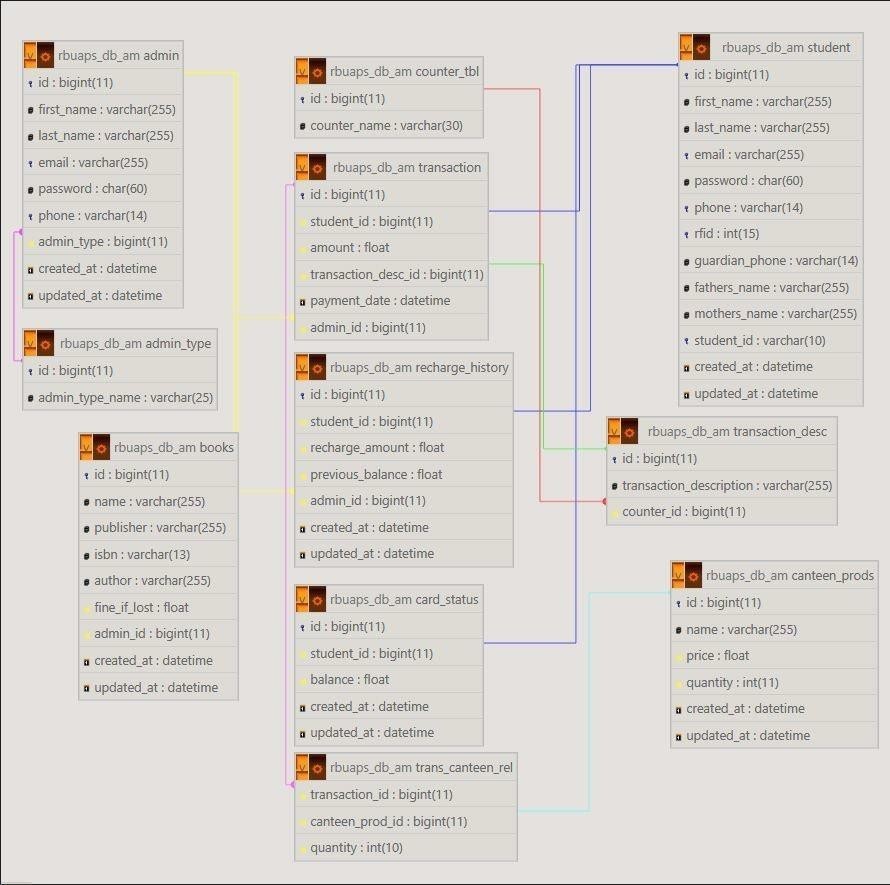}
    \caption{ER Diagram of the RFID-Based Payment System.}
    \label{fig:er_diagram}
\end{figure}

\subsection{Summary}
This methodology ensures a robust and scalable implementation of the RFID-based payment system. The integration of RFID technology with campus services streamlines operations, improves security, and enhances the user experience, making it a viable solution for educational institutions.

\section{Results and Discussion}

The implementation of the RFID-based campus-wide payment system demonstrates a significant step towards creating a cashless, efficient, and secure financial ecosystem for educational institutions. The integration of RFID technology with a Raspberry Pi backend server and a dynamic website has achieved the primary objectives of this study. The key results and their implications are discussed below.

\subsection{System Functionality and Performance}
The proposed system utilizes RFID technology integrated with a Raspberry Pi, serving as a cost-effective and compact backend server. The system successfully handles the following functionalities:
\begin{itemize}
    \item \textbf{Student Payment Management:} Students use RFID cards to pay admission fees, tuition fees, canteen bills, and library fines. The transactions are processed in real-time, with amounts deducted from the student's account balance and recorded in the database.
    \item \textbf{Dynamic Website Interface:} A user-friendly website was developed to allow:
    \begin{itemize}
        \item Students to view transaction history, including canteen and recharge records.
        \item Admins to perform various tasks such as registering new students, adding suppliers, and managing café or accounts-related activities. Admins can also retrieve and update student records efficiently.
    \end{itemize}
    \item \textbf{Backend Server and Database:} The backend server processes payments and maintains a robust database storing user details, transaction history, and account balances. The integration ensures seamless communication between the RFID reader and the server.
\end{itemize}

\subsection{Cost-Effectiveness of Raspberry Pi}
Using Raspberry Pi as the central server eliminates the need for traditional desktop computers, significantly reducing setup costs. Raspberry Pi operates as a minicomputer, offering sufficient processing power to handle the requirements of this system. This approach not only minimizes hardware expenses but also promotes scalability. If Raspberry Pi production is localized, the system's cost can become negligible, making it highly accessible for educational institutions.

\subsection{Real-Time Transaction Processing}
The system's ability to process transactions in real time was a critical success factor. When a student uses their RFID card for payments, the RFID reader immediately sends the unique identifier to the server. The server then:
\begin{itemize}
    \item Validates the card details.
    \item Deducts the appropriate amount from the student's account balance.
    \item Updates the transaction history in the database.
\end{itemize}
This ensures a seamless and efficient payment process for both students and administrators, reducing delays and errors commonly associated with manual or cash-based systems.

\subsection{Enhanced User Experience}
The website interface and system design focused on user convenience:
\begin{itemize}
    \item \textbf{Students:} Students can log in to the website to view their transaction and recharge history, offering complete transparency and control over their financial activities.
    \item \textbf{Admins:} Admins can easily manage student data, transaction records, and administrative tasks such as adding new suppliers or creating specific admin roles for activities like café management.
\end{itemize}

\subsection{Challenges and Limitations}
While the system meets its objectives, certain challenges were identified:
\begin{itemize}
    \item \textbf{Dependency on RFID Cards:} The system's reliance on RFID cards means that card loss or damage could disrupt user access. A mechanism for quick card replacement is essential.
    \item \textbf{Scalability:} As the number of students increases, the system may face challenges in terms of database performance and transaction processing speed. Future work could focus on optimizing the database and server infrastructure for scalability.
    \item \textbf{Local Production Feasibility:} While Raspberry Pi offers a cost-effective solution, the feasibility of local production to further reduce costs requires additional exploration.
\end{itemize}

\subsection{Discussion and Future Work}
The implementation demonstrates that RFID technology, when paired with a cost-efficient Raspberry Pi backend, is a viable solution for campus-wide payment systems. The system effectively eliminates the need for cash-based transactions, enhancing convenience and security for students and administrators alike.

Future enhancements could include:
\begin{itemize}
    \item Integrating mobile application support for additional accessibility.
    \item Employing advanced security measures, such as biometric verification, to complement RFID-based authentication.
    \item Expanding the system's scalability to accommodate a larger user base without compromising performance.
\end{itemize}

Overall, this work underscores the potential of RFID technology to transform traditional payment methods in educational institutions while maintaining a focus on cost-efficiency, user-friendliness, and scalability.

\section{Conclusion}

The RFID-based campus-wide payment system offers a modern and efficient approach to managing financial transactions in educational institutions. By integrating RFID technology, this system streamlines payment processes, reduces transaction time, and enhances security. Students can conveniently make payments for various campus services, such as cafeteria bills, library fines, and tuition fees, by simply using their RFID-enabled cards. This ensures a hassle-free experience and eliminates the need for carrying cash, thereby promoting a secure and cashless campus environment.

Additionally, the system's centralized payment management simplifies the overall financial operation for administrators, reducing the complexities associated with cash handling and improving accountability. The adoption of this system not only enhances convenience for students but also contributes to a safer and more transparent campus financial ecosystem.

However, the successful implementation of an RFID-based payment system requires careful planning, proper infrastructure setup, and comprehensive user training. Addressing these prerequisites ensures the system's efficiency and usability, leading to a positive impact on the campus experience.

In conclusion, the RFID-based campus-wide payment system provides a seamless and effective payment solution, significantly improving the quality of campus life for students and administrators alike. This work demonstrates the potential of RFID technology to transform traditional payment methods into a more modern, secure, and user-friendly solution for educational institutions.

\end{document}